\documentclass[10pt,aps,prl,twocolumn,superscriptaddress]{revtex4-1}

\usepackage[utf8]{inputenc}
\usepackage{hyperref}
\usepackage{epsfig}
\usepackage{graphicx}
\usepackage{amsmath}
\usepackage{amssymb}
\usepackage{color}

\graphicspath{{Images/}}

\begin{document}

\title{
Video-rate computational heterodyne holography 
}

\author{Antoine Dillée}
\author{Romain Cancilliere}
\affiliation{
Langevin Institute. Centre National de la Recherche Scientifique (CNRS) UMR 7587, Institut National de la Sant\'e et de la Recherche M\'edicale (INSERM) U 979, Universit\'e Pierre et Marie Curie (UPMC), Universit\'e Paris Diderot. Ecole Sup\'erieure de Physique et de Chimie Industrielles (ESPCI) - 1 rue Jussieu. 73205 Paris, France
}
\author{Fernando Lopes}
\affiliation{
Geomagnetism and Paleomagnetism, Institut de Physique du Globe de Paris, Universit\'e Paris Diderot, Sorbonne Paris Cit\'e, 1 rue Jussieu, Paris, France
}
\author{Michael Atlan}
\affiliation{
Langevin Institute. Centre National de la Recherche Scientifique (CNRS) UMR 7587, Institut National de la Sant\'e et de la Recherche M\'edicale (INSERM) U 979, Universit\'e Pierre et Marie Curie (UPMC), Universit\'e Paris Diderot. Ecole Sup\'erieure de Physique et de Chimie Industrielles (ESPCI) - 1 rue Jussieu. 73205 Paris, France
}

\date{\today}
\begin{abstract}
We present a versatile computational image rendering software of optically-acquired holograms. The reported software can process 4 Megapixel 8-bit raw frames from a sensor array acquired at a sustained rate of 80 Hz. Video-rate image rendering is achieved by streamline image processing with commodity computer graphics hardware. For time-averaged holograms acquired in off-axis optical configuration with a frequency-shifted reference beam, wide-field imaging of one tunable spectral component is permitted. This software is validated by phase-stepped hologram rendering, and non-contact monitoring of surface acoustic waves by single and dual sideband hologram rendering. It demonstrates the suitability of holography for video-rate computational laser Doppler imaging in heterodyne optical configuration.
\end{abstract}
               
\maketitle


Heterodyne holography, a variant of time-averaged holography~\cite{Powell1965, PicartLeval2003} makes use of a frequency-shifted reference beam~\cite{Aleksoff1971} to downconvert radiofrequency (RF) optical fluctuation spectra to the sensor bandwidth. It is a good candidate for single-frequency mechanical vibrations mapping of nanometric~\cite{UedaMiida1976, PsotaLedl2012, VerrierAtlan2013, BrunoLaudereau2013} to micrometric~\cite{JoudVerpillat2009} amplitudes. Real-time holographic imaging was a technological issue, hindering real-world usability of holographic detection schemes, until the advent of graphics processing units (GPUs). Computational rendering techniques of optical holograms on GPUs~\cite{ShimobabaSato2008, Ahrenberg2009} have led to real-time qualitative holographic vibration imaging of nanometric out-of-plane vibrations at narrow~\cite{SamsonVerpillat2011} and wider~\cite{SamsonAtlan2013} detection bandwidths. GPU image rendering has enabled fast auto-focusing of megapixel-resolution holograms~\cite{Dogar2013}, speckle diversity rendering~\cite{TrujilloGarciaSucerquia2013}, three-dimensional microscopic scenes rendering~\cite{Orzo2010, Mudanyali2010, Cheong2010, ShimobabaMasuda2010, VerpillatJoudDesbiolles2011, HobsonReid2013}. For time-consuming holographic image rendering, in particular when real-time performance is required, hardware acceleration of computations should be considered~\cite{GaoKemao2012, Reid2012}.\\

In this letter, a video-rate computational holographic image rendering software is presented. We demonstrate processing of 4 Megapixel interferogram recordings at a throughput of 320 Mega bytes per second with commodity graphics hardware. At the difference of the available libraries~\cite{ShimobabaIto2008, ShimobabaWeng2012}, the software is readily usable to record holographic videos. Experimental demonstrations of phase-stepped video holography and quantitative remote monitoring of optical sideband holograms are reported.\\


%
\begin{figure}[b]
\centering
\includegraphics[width = 6.0 cm]{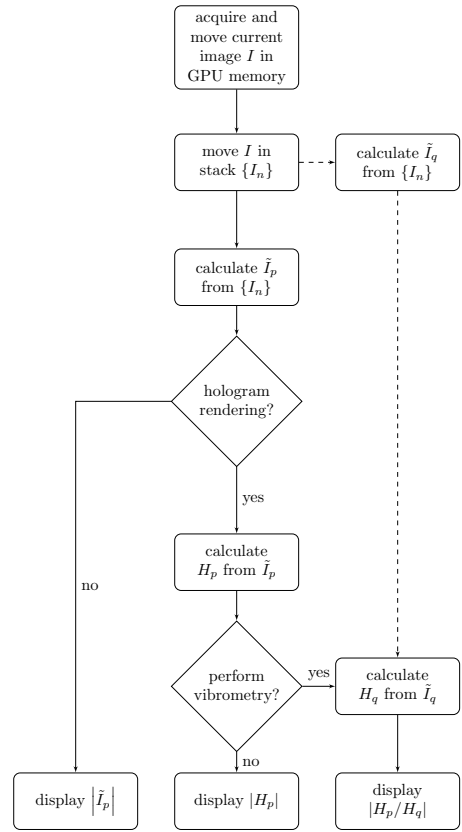}
\caption{Flow chart of holographic image acquisition, rendering and display. Raw interferograms are recorded by the acquisition thread. Temporal and spatial signal demodulation are performed by the hologram rendering thread. Image display is performed by a separate thread.}
\label{fig_FlowChart}
\end{figure}

The aim of the developed software is to compute holograms on a GPU from interferograms recorded with high throughput cameras. It is essential that the acquisition of the raw video stream from the camera does not miss any image, in order to ensure the consistency of the temporal demodulation. No other tasks should therefore disrupt video capture. The software has four main threads dedicated to (i) image acquisition, (ii) holographic image rendering, (iii) image display, and (iv) image saving. The three first threads run continuously, while the user can choose between image display and recording. The software matches real-time constraints. Temporal and spatial demodulation calculations are handled by the GPU, by an algorithm elaborated with Microsoft Visual C++ 2012 and NVIDIA's Compute Unified Device Architecture (CUDA) software development kit 5.5, on single precision floating point arrays. A flowchart of its algorithm is reported in Fig.~\ref{fig_FlowChart}.\\


The main image acquisition thread saves the interferogram $I(t)$ recorded at time $t$ in the random-access memory of the computer. For the sake of performance, non-paged memory is preferred to virtual memory. However, some cameras did not support saving images into non-paged memory. Hence the software checks if the camera supports non-paged memory and whether to use it to save the image. Some cameras choose their own buffer in contrast to others which accept forced recording to any address in memory. \\


The hologram rendering thread constitutes the core of the software; it performs both temporal and spatial signal demodulation. Before treatment, the GPU processors are initialized for parallel processing. The program automatically adapts to the onboard graphics card in the computer and the resolution of the image to be processed through the use of CUDA functions. The newest recorded interferogram $I(t)$ at time $t$ is added to a "First In, First Out" (FIFO) stack of images $\{ I_n \equiv I(t - n / \nu_{\rm S} )\}$ from which the oldest image is deleted, so that the stack is always refreshed with the last $N$ frames recorded. To facilitate image processing, raw images are zero-padded to $2^M\times 2^M$ elements arrays, where $2^M$ is the nearest integer power of 2 above the larger lateral dimension of the camera. The $N$ last recorded interferograms $\{ I_n \}$, where $n \in \{0, \ldots, N-1 \}$ are temporally-demodulated numerically by short-time discrete Fourier transformation (DFT)~\cite{SamsonAtlan2013}. The $p$-th component of the DFT takes the form
\begin{equation}\label{eq_interferogram}
\tilde{I}_p = \sum_{n=0}^{N-1} I_{n} \exp \left(- 2 i \pi np / N \right)
\end{equation}
where $p \in \{0, \ldots, N-1 \}$. The calculated array is a matrix of $2^M \times 2^M$ complex scalar values $\tilde{I}_p[m,n]$, with pixel indexes $(m,n) \in \{1, \ldots, 2^M \} \times \{1, \ldots, 2^M \}$. This array carries the optical field distribution in the sensor plane, which is the input of the spatial demodulation process. Spatial signal demodulation - or holographic image rendering - from temporally-demodulated interferograms involves numerical Fourier transforms. The spatial Fourier transform ${\cal F}$ and its inverse ${\cal F}^{-1}$ are computed with Fast Fourier Transform (FFT) algorithms by the function \emph{cufft2d()}. Several methods can be used for hologram reconstruction~\cite{Schnars2002, KimYuMann2006, PicartLeval2008, VerrierAtlan2011}, among which we implemented the most basic ones : the Fresnel (1-FFT), and the angular spectrum propagation approaches (2- and 3-FFT).\\

Under the paraxial approximation, for optical systems of low numerical aperture, the propagation integral of the optical field can be written as a Fresnel transform. Spatially-demodulated holograms by the 1-FFT rendering method take the form
\begin{equation}\label{eq_1FFT_Propagation}
H_p =  {\cal F} \{ \tilde{I}_p \times h \}
\end{equation}
where $h[m,n]$ is the discretized free-space propagation impulse response
\begin{equation}\label{eq_h}
h = \exp \left[i k (x^2 + y^2) / (2z)\right]
\end{equation}
where $x[m,n] = m d$ and $y[m,n] = n d$, $k = 2\pi/\lambda$, with $(m,n) \in \{1, \ldots, 2^M \} \times \{1, \ldots, 2^M \}$. The parameter $z$ corresponds to the sensor-to-object distance for a flat LO wavefront and in the absence of lens in the object path. Since only the magnitude of the optical field is displayed, additional phase factors are not taken into account in the expression of $h$. This rendering method is suited for macroscopic imaging, enabling the reconstruction of holograms of extended objects located far away from the sensor. The propagation integral can also be viewed as the propagation of the angular spectrum of the optical radiation. It can be derived from the Helmholtz equation and is suitable for computing reverse diffraction in a high numerical aperture optical system, in which use of Fresnel diffraction is not suitable. In convolution approaches, spatial demodulation can be expressed as a spatial convolution product, computed with Fourier transforms
\begin{equation}
H_p = {\cal F}^{-1} \{ {\cal F} \{ \tilde{I}_p \} \times \exp  ( i z \sqrt{ k^2 - k_x^2 - k_y^2 } ) \}
\label{eq_2FFT_Propagation}
\end{equation}
where $k_x[m,n] = 2 \pi m / (2^M d)$, and $k_y[m,n] = 2 \pi n / (2^M d)$, with $(m,n) \in \{1, \ldots, 2^M \} \times \{1, \ldots, 2^M \}$. Eq.~\ref{eq_2FFT_Propagation} constitutes the algorithm of 2-FFT rendering. In the case of 3-FFT rendering, the angular spectrum kernel is computed as the FFT of the free-space propagation impulse response $h$. The spatially-demodulated image takes the form
\begin{equation}
H_p = {\cal F}^{-1} \{ {\cal F} \{ \tilde{I}_p \} \times {\cal F} \{ h \} \}
\label{eq_3FFT_Propagation}
\end{equation}
Angular spectrum propagation methods are typically used for microscopic imaging, when dealing with small objects located near the sensor.\\

 
The display thread is designed to view a sustained video stream of processed holograms. To this aim, the graphics card of the computer is set to profit by having recourse to an OpenGL control. The user has the ability to zoom the display part of the image by drawing a selection rectangle. The refreshment rate of the display is set to 25 Hz. Three kinds of images can be displayed (Fig.~\ref{fig_FlowChart}) : the magnitude of temporally-demodulated interferograms $|\tilde{I}_p|$, the magnitude of holograms $|H_p|$, and the magnitude of the ratio of two holograms $|H_p/H_q|$, where $p$ and $q$ $\in \{ 0, \ldots, N-1\}$ are demodulation frequency indexes. For example, to view a standard video of sequentially grabbed frames, $|\tilde{I}_0|$ should be displayed; the parameters of the temporal demodulation should be set to $N = 1$ and $p = 0$. To achieve phase-stepped interferometry~\cite{Creath1985} by computing the modulated component of the interferogram at one fourth of the sampling frequency $\nu_{\rm S}/4$, $|\tilde{I}_1|$ should be displayed with $N = 4$.  To perform 4-phase holography, $|H_1|$ should be displayed with $N = 4$. To perform  the ratio of two holograms modulated at $\pm \omega_{\rm S} /4$~\cite{VerrierAtlan2013}, $|H_1/H_3|$ should be displayed with $N = 4$.\\


The image saving thread saves a series of raw ($\{ I_n \}$) or processed ($\{ |\tilde{I}_p |\}$, $\{  |H_p| \}$, or  $\{  |H_p/H_q| \}$) images. A series of images is written in a binary file (.bin). The user can select the recording directory and the number of images to record. A log file containing input parameters (distance parameter $z$, wavelength $\lambda$, spatial rendering method used, image size and bit depth) is created. Depending on their original format, data are coded on 8-bit or 16-bit per pixel, with byte order convention ”little endian” for 16-bit quantization.\\


%
\begin{figure}[]
\centering
\includegraphics[width = 8 cm]{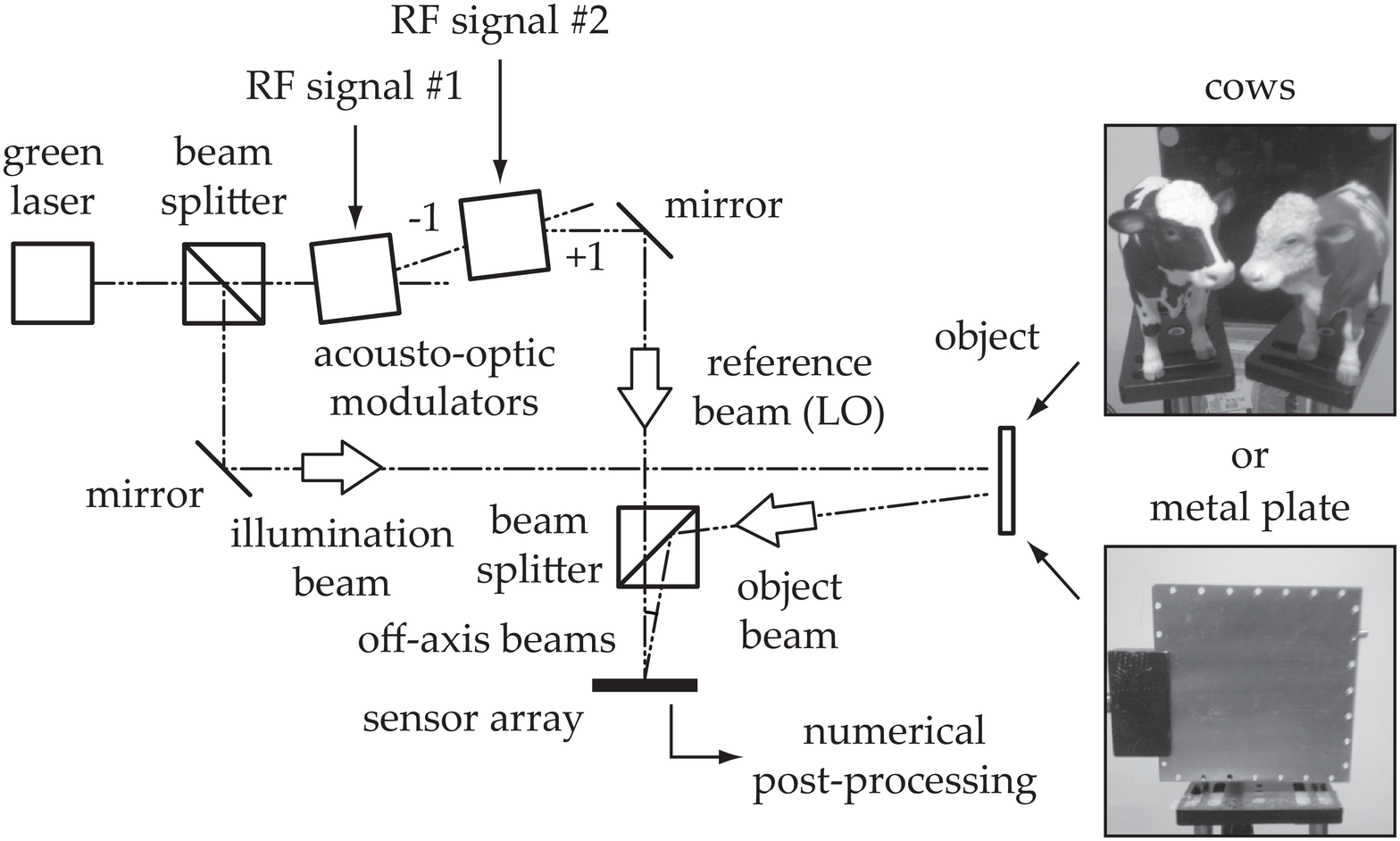}
\caption{Optical arrangement. The main laser beam is split into two channels, forming a Mach-Zehnder interferometer. The probe beam is backscattered by an object, yielding the optical field $E$. In the reference channel, the optical local oscillator field $E_{\rm LO}$ is frequency-shifted by two acousto-optic modulators, driven by phase-locked frequency synthesizers, from which alternate diffraction orders ($\pm1$) are selected. Holographic images are computed numerically from interferometric measurements of the diffracted object beam beating against the local oscillator beam.}
\label{fig_Setup}
\end{figure}

Optical recording of interferograms is performed with the setup sketched in Fig.~\ref{fig_Setup}. It is an off-axis, frequency-shifted Mach-Zehnder interferometer used for optical heterodyne detection of an object field $E$ beating against a separate local oscillator (LO) field $E_{\rm LO}$, on a sensor array, in reflective geometry~\cite{AtlanGross2006, SamsonVerpillat2011, VerrierAtlan2013}. The main optical field is provided by a 100 mW, single-mode laser (wavelength $\lambda = 532$ nm, optical frequency $\nu_{\rm L} = \omega_{\rm L}/(2 \pi) = 5.6 \times 10^{14} \, \rm Hz$, Oxxius SLIM 532). The LO beam is diffracted by two acousto-optic modulators (AA-electronics, MT80-A1.5-VIS) driven by continuous-wave radiofrequency signals~\cite{SamsonVerpillat2011, JoudVerpillat2009, VerrierAtlan2013, BrunoLaudereau2013}. The observed objects are shined over $\sim 10 \, {\rm mm} \times 10 \, {\rm mm}$ to $\sim 200 \, {\rm mm} \times 200 \, {\rm mm}$ with $\sim$ 50 mW of light power. When shaken with a piezo-electric actuator (PZT, Thorlabs AE0505D08F), driven sinusoidally, their nanometric vibration provoke a local sinusoidal phase modulation $\phi$ of the backscattered optical field $E$. This modulation results in the apparition of optical sidebands at the harmonics of the modulation frequency~\cite{Powell1965, Aleksoff1971, JoudVerpillat2009, SamsonVerpillat2011}. For the reported experiments, interference patterns are measured with a Ximea MQ042MG-CM camera (pixel size $d = 5.5 \, \mu \rm m$), run in external trigger mode at a frame rate of $\nu_{\rm S} = \omega_{\rm S} / (2 \pi) = 80 \, \rm Hz$, at 8 bit/pixel quantization. Each raw interferogram of $2048 \times 2048$ pixels is digitally acquired at time $t$ is noted $I(t) =  \left| E(t) + E_{\rm LO}(t) \right| ^2$.\\



%
\begin{figure}[]
\centering
\includegraphics[width = 8 cm]{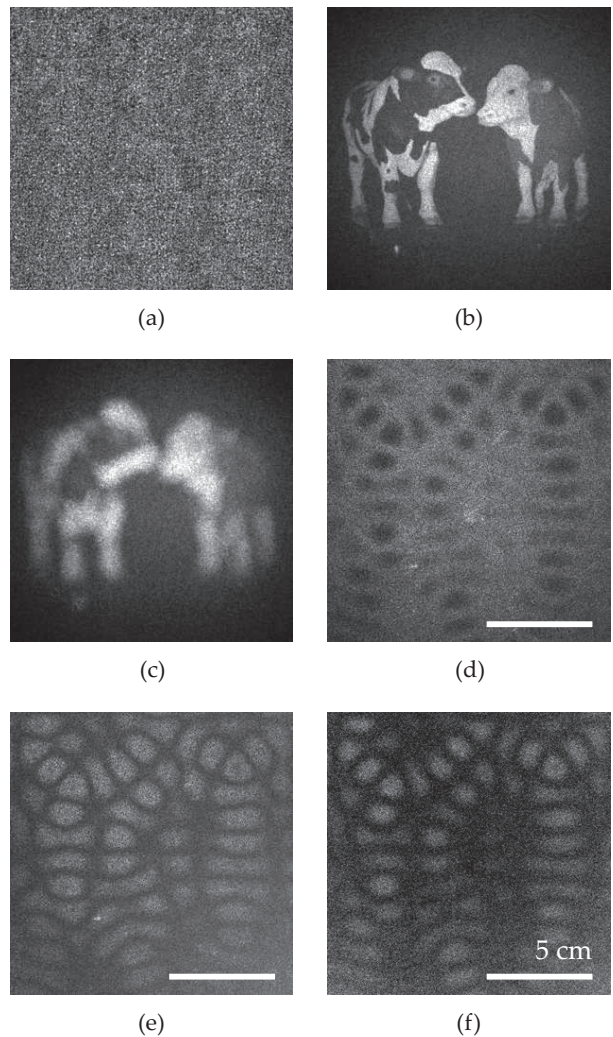}
\caption{Images of the cows : $|\tilde{I}_1 |$ (a), $|H_1|$ at $z = 0.47\, \rm m$ (b), $|H_1|$ at $z = 0.39\, \rm m$ (c). Contour modes of the plate vibrating at 21.5 kHz : $|H_1|$ (d), $|H_3|$ (e), and $|H_3/H_1|$ (f).}
\label{fig_Holograms}
\end{figure}

In a first experiment, we  performed phase-stepped video holography of a static object (plastic cows reported in Fig.~\ref{fig_Setup}). The optical probe field $E$  is a monochromatic scalar wave of temporal dependence $\sim \exp \left( i \omega _{\rm L} t \right)$. The temporal dependence $\sim \exp \left( i \omega_{\rm L} t + i \Delta \omega t \right)$ of the optical LO field $E_{\rm LO}$ provokes a modulation of the cross-beating contributions in the interference pattern $I(t)$ at the angular frequency $\Delta \omega$. This modulation can be used for accurate~\cite{AtlanGross2007} phase-stepped measurements. For instance, if the frequency detuning is set to a quarter of the camera frame rate $\Delta \omega = \omega_{\rm S}/4$, the arrays $\tilde{I}_1$ and $H_1$ are the 4-phase demodulated interferogram (Fig.~\ref{fig_Holograms}(a)) and hologram (Fig.~\ref{fig_Holograms}(b)), respectively. Their amplitude, proportional to the object field magnitude, is refreshed at video rate. The distance parameter $z$ was changed during hologram rendering between the images reported in Fig.~\ref{fig_Holograms}(b), and Fig.~\ref{fig_Holograms}(c).\\


In a second experiment, we performed quantitative imaging of nanometric optical pathlength modulations, by using a dual LO signal input in one of the acousto-optic modulators~\cite{VerrierAtlan2013}. This experiment was previously reported, with offline post-processing~\cite{BrunoLaudereau2013}. We assessed a thin metal plate's out-of-plane vibration modes. The structure was excited sinusoidally at the angular frequency $\omega$. The resulting transverse motion $z(t)$ at a given point of the surface of the plate, is a sine wave at the angular frequency $\omega$, with local modulation depth $z_0$. The local sinusoidal phase modulation of the backscattered optical field $E$ takes the form $\phi(t) = 2 k z_0 \sin (\omega t)$. To assess local vibration amplitudes, the hologram $|H_1|$ was tuned on the non-shifted light component (Fig.~\ref{fig_Holograms}(d)). The hologram $|H_3|$ (Fig.~\ref{fig_Holograms}(e)) was tuned on the first optical modulation sideband at $\omega_{\rm L} + \omega$. Narrowband detection of the local out-of plane vibration amplitude was realized by forming the quantity $|H_3/H_1| \approx k z_0$, reported in Fig.~\ref{fig_Holograms}(f).\\


In conclusion, we have designed a video-rate computational narrowband laser Doppler imaging scheme with optically-acquired heterodyne holograms and stream processing on commodity graphics hardware. With the reported hardware configuration, sustained sampling of interferograms at a throughput of 320 Mega bytes per second was achieved. Real-time image rendering of holograms involving temporal signal demodulation by short-time discrete Fourier transform and spatial demodulation methods involving fast Fourier transforms was achieved. To this end, we created a robust and versatile software running on Microsoft Windows 7 64-bit. In single LO regime, video-rate computational holography could be used to monitor Doppler frequency lines due to flows~\cite{AtlanGrossLeng2006}, or thermal motion~\cite{AtlanDesbiolles2010}. In dual LO regime, video-rate computational holography could be used for investigating low optical pathlength modulation effects at known frequencies, such as photothermal modulation~\cite{AbsilTessierGross2010}, sinusoidal motion of small particles~\cite{AbsilTessierFournier2009}, and surface acoustic waves~\cite{BrunoLaudereau2013}. To enable wideband detection of optical fluctuations above the kilohertz range, one of the remaining technical challenges of computational holography for Doppler imaging with optical sensor arrays lies less in the processing power of GPUs than in the camera streaming throughput.\\



We gratefully acknowledge support from Fondation Pierre-Gilles de Gennes (FPGG014), Agence Nationale de la Recherche (ANR-09-JCJC-0113, ANR-11-EMMA-046), r\'egion \^Ile-de-France (C'Nano, AIMA).\\


\end{document}